\documentclass[final,5p,times,number]{elsarticle}
\pdfoutput=1 %pour imposer pdflatex

\usepackage[utf8]{inputenc}
\usepackage{graphicx}
\usepackage{amssymb}
\usepackage{amsmath}
\usepackage{subcaption}
\usepackage{url}

\usepackage{lineno}
\usepackage{setspace}
\usepackage{comment}
\usepackage{textcomp}
\newcommand{\textapprox}{\raisebox{0.5ex}{\texttildelow}}

\usepackage[load-configurations = abbreviations]{siunitx}
\usepackage{color}
\newcommand{\revised}{ }

\biboptions{sort&compress}
\journal{Acta Materialia}
\begin{document}

\begin{frontmatter}
\begin{singlespace}
\title{Metrology of small particles and solute clusters by atom probe tomography}

\author[simap1]{Frédéric De Geuser}
\ead{frederic.de-geuser@simap.grenoble-inp.fr}
\author[mpie,imperial]{Baptiste Gault}
\ead{b.gault@mpie.de}

\address[simap1]{Univ. Grenoble Alpes, CNRS, Grenoble INP, SIMaP, F-38000 Grenoble, France}
\address[mpie]{Max-Planck Institut für Eisenforschung, Max-Planck-Straße 1,
D-40237 Düsseldorf, Germany}
\address[imperial]{Department of Materials, Royal School of Mines, Imperial College, Prince Consort Road, London, SW7 2BP, UK}

\begin{abstract}
Atom probe tomography (APT) is routinely used for analyzing property-enhancing particles in the nanometer-size range and below, and plays a prominent role in the analysis of solute clusters. However, the question of how well these small particles are measured has never been addressed because of a lack of a reliable benchmark. Here, to address this critical gap, we use an approach that allows direct comparison of APT and small-angle (X-Ray) scattering (SA(X)S) performed on the same material. We introduce the notion of an effective spatial resolution for the analysis of particles, which, importantly in this context, is very different than the technique's inherent spatial resolution. This effective resolution is highly specific to the system being considered, as well as the analysis conditions. There is no hard limit below which the technique will fail, but particles with a radius of order of \textapprox $2\sigma=\SI{1}{\nm}$, i.e. \textapprox\num{250} atoms cannot be accurately measured, even though the particles are detected.  This thorough metrological assessment of APT in the analysis of particles allows us to discuss the pulse spread function of the technique and the physics underpinning its limits. We conclude that great care should be taken when analysing solute clusters by APT, in particular when reporting particle size and composition. 
\end{abstract}

\end{singlespace}

\end{frontmatter}

\section{Introduction}
Atom probe tomography (APT) has progressively become prominent as a tool for analyzing the composition of microstructural features, i.e.  precipitates, in the nanometer-size range and below \citep{Marquis2013,Devaraj2018}. 
This is particularly true for solute clusters  \citep{Dumitraschkewitz2018}.  
The detailed analysis of the composition of these precipitates and clusters is of high relevance to studies of the early stages of nucleation of new phases, which control both the physical properties of materials and the further evolution of the microstructure. If APT is often presented as a microscopy and microanalysis technique, it is primarily a mass spectrometry technique \citep{Muller1968} with a capacity to map the composition in three-dimensions in a volume of solid material in the range of  \SI{100 x 100 x 500}{\nm} \citep{Blavette1993b}.  

{\revised 
There is a significant gap in the literature regarding the size of the smallest microstructural object that can be precisely analysed. In a sense, this size could be seen as an effective spatial resolution, below which individual objects cannot neither be accurately nor precisely characterized. Yet this concept is difficult to define because: (i) it is likely highly dependent on the system being studied; (ii) the nature of the very small microstructural objects or particles, i.e. solute clusters, is ill-defined at the atomic scale. There are still debates as to how to define a solute cluster and whether it is possible to distinguish between a random fluctuation and a ``real'' cluster.

The presence of such clusters is usually assessed by comparing the experimental data to an equivalent dataset in which the mass-to-charge ratio has been randomly swapped between all the atoms present in a reconstruction \citep{Moody2007, Stephenson2007} or a statistically random distribution \citep{Moody2008}. The comparison to a random distribution is necessary to characterise a population of clusters \citep{Cerezo2007b, Stephenson2007}, but may not be sufficient. Cluster-finding techniques typically rely on a fixed distance threshold, often referred to as $d_{max}$, below which atoms are considered to belong to a cluster \citep{marquis_applications_2010, Dumitraschkewitz2018}. Various criteria have been proposed to define $d_{max}$ based on the comparison to random \citep{Hyde2011}, yet there is no widely spread approach that has been shown to lead to consistent results across the community \citep{Marquis2016, Dong2019}. Alternatives have been proposed, using radial distribution functions \citep{Marquis2002,DeGeuser2006}, and their quantitative analysis \citep{couturier_direct_2016, Zhao2018a}}.

{\revised How to define APT's effective spatial resolution? For a microscope, an infinitely small point source observed in the image plane through the entire imaging system usually appears distorted. The shape and size of the image of the point source can be assumed to be the response of the microscope to an impulse, and is referred to as the point spread function (PSF). The PSF can be seen as the base unit of an image and it imposes a size limit: an object smaller than its size will appear having the size of the PSF}. 

Importantly, the PSF is not the technique's inherent spatial resolution or resolution limit. These two have only been assessed based on the analysis of known crystallographic parameters of the sampled material and in the case of pure materials \cite{vurpillot_structural_2001, gault_spatial_2010, Kelly2009, geiserSpatialDistributionMaps2007}. The spatial resolution is known to be non-isotropic, and better in depth than laterally. 

Regarding the lateral resolution, local variations of the electrostatic field cause deflections in the ions trajectories in the early stage of their flight \citep{Vurpillot2015}. 
This combines with the possibility that atoms have a rolling motion on their nearest neighbours at the surface before departure to cause a strong uncertainty on the atomic positions upon reconstruction \cite{Suchorski1994}. Both aspects have a random component and are driven by the local arrangement of the atoms at the surface and are hence nearly impossible to predict and correct. The lateral resolution was shown to vary with the respective atomic packing on the terraces corresponding to various sets of crystallographic planes intersecting the specimen's surface \citep{gault_spatial_2010-1}.

Regarding the depth resolution, it is limited in part by the same aberrations but also by the reconstruction protocols. The current algorithm used to build the tomographic reconstruction makes use of the sequence in which the ions are collected by the position-sensitive detector to assign a depth coordinate to each reconstructed ion \citep{vurpillot_reconstructing_2013}. The depth resolution varies significantly across the field of view of a specimen, based on the relative field evaporation behaviour of different crystallographic facets, as well as with the experimental conditions and the material under investigation \citep{Cadel2009, gault_spatial_2010-1, Gault2011}. Besides, the concept of depth resolution itself is ambiguous since it relates to a direction normal to the surface of the specimen, rather than to a fixed direction with respect to the reconstructed volume \citep{gault_spatial_2010-1}. The high value of the depth resolution can only be reached within a sub-volume with a cross-section of only \textapprox\  \SI{2x2}{\nm} to \SI{5x5}{\nm} positioned close to crystallographic poles \citep{gault_spatial_2010-1}, and varies from pole to pole. 

The question of the spatial resolution has thus been essentially addressed for near-ideal cases and in subvolumes where the resolutions is best, i.e. close to poles where one or more sets of atomic planes were imaged. This can be ascribed to the fact that, first, atomic planes were the only reliable benchmarks that were available, with the notable exception of \cite{Shimizu2011} who used isotopic multilayers, and, second, that the community was in search of a ultimate spatial resolution value.

These studies were performed mostly on pure materials and leave unanswered the critical question of the accuracy of the technique in the analysis of actual property-enhancing microstructural features that exhibit a different composition, and potentially structure, compared to the surrounding matrix. In the case of secondary phases, examples are sparse but revealing. 
{\revised Araullo-Peter studied T$_1$ precipitates in an Al-matrix. For a single microstructural feature of interest, they showed that the decrease in spatial resolution caused by local changes in the field evaporation process can conceal obvious interfacial segregation \citep{Araullo-Peters2014a}. This work disproved previous claims by Gault et al. \citep{gault_atom_2011}}.

To our knowledge, the question of the effective spatial resolution for APT in the analysis of particles has never been addressed. This effective spatial resolution, which relates directly to the PSF, results from a combination of lateral and in-depth resolution applied to the study of clusters or precipitates, and relevant to the entire data set.

Here, we discuss an approach to estimate this effective spatial resolution in the analysis of particles with size below \SI{10}{\nm}, including solute clusters.  
We exploit a framework which allows direct comparison of APT and small-angle scattering (SAS). 
We report results with X-Rays (SAXS) performed on the same material as the APT analysis. 
SAXS allows to detect compositional fluctuations on the smallest scale and does not suffer from the same artefacts as APT. 
{\revised Comparing between} techniques for a range of Al-alloys and steels, we quantify the range of feature size, i.e. radius of a spherical particle, for which APT fails to register the features of interest with relevant accuracy, and discuss the reasons underpinning why APT does not seem able to report sizes of particles with a radius below \textapprox \SI{1}{nm}.  
To put our results into a broader perspective, we compare them to values from the recent literature, including studies that combined small-angle neutron scattering (SANS) or SAXS together with APT, highlighting the general character of our observations.

\section{Framework for APT / SAXS comparison}
The approach we propose here makes use of radial distribution functions (RDF) calculated from within the reconstructed APT data \citep{Marquis2002, Sudbrack2006, haley_influence_2009}. 
The neighborhood of each ion of a specific species is interrogated, and an average composition of each species as a function of the radial distance to each ion is established. 
De Geuser and co-workers described how to process RDF from atom probe data using the formalism typically used to process small-angle scattering data \cite{couturier_direct_2016,Zhao2018a}. 
They introduced an $i-j$ pair correlation function (PCF) \citep{DeGeuser2006}, $\gamma_{i-j}(r)$, between element $i$ and element $j$ by normalization and scaling of the RDF:
\begin{equation}
\gamma_{i-j}(r)=C_i C_{i-j}(r)-C_iC_j
\end{equation}
where $C_{i-j}(r)$ is the average local composition of $j$ at a distance $r$ of atoms of element $i$, as typically obtained from any RDF computation software. 
$C_i$ and $C_j$ are respectively the average composition of element $i$ and $j$ in the considered volume. 
In particular, when $i=j$, we have:
\begin{equation}
\gamma_{i-i}(r)=C_i C_{i-i}(r)-C_i^2
\end{equation}

Within this definition, the value of the pair correlation function becomes 0 at large $r$ and its value at $r=0$ corresponds to the mean square fluctuation which, in a two-phase system, i.e. precipitates and matrix, can be written:
\begin{equation}
\gamma_{i-i}(0)=\overline{\Delta C_i^2}=(C_p-C)(C-C_m)\label{mean_sq}
\end{equation}
where we have dropped the $i$ indices for the right part of the equation, and where $C$, $C_m$ and $C_p$ are the average composition, the matrix composition and the precipitates composition respectively.

\begin{figure}
    \centering
    \includegraphics[width=8.5cm]{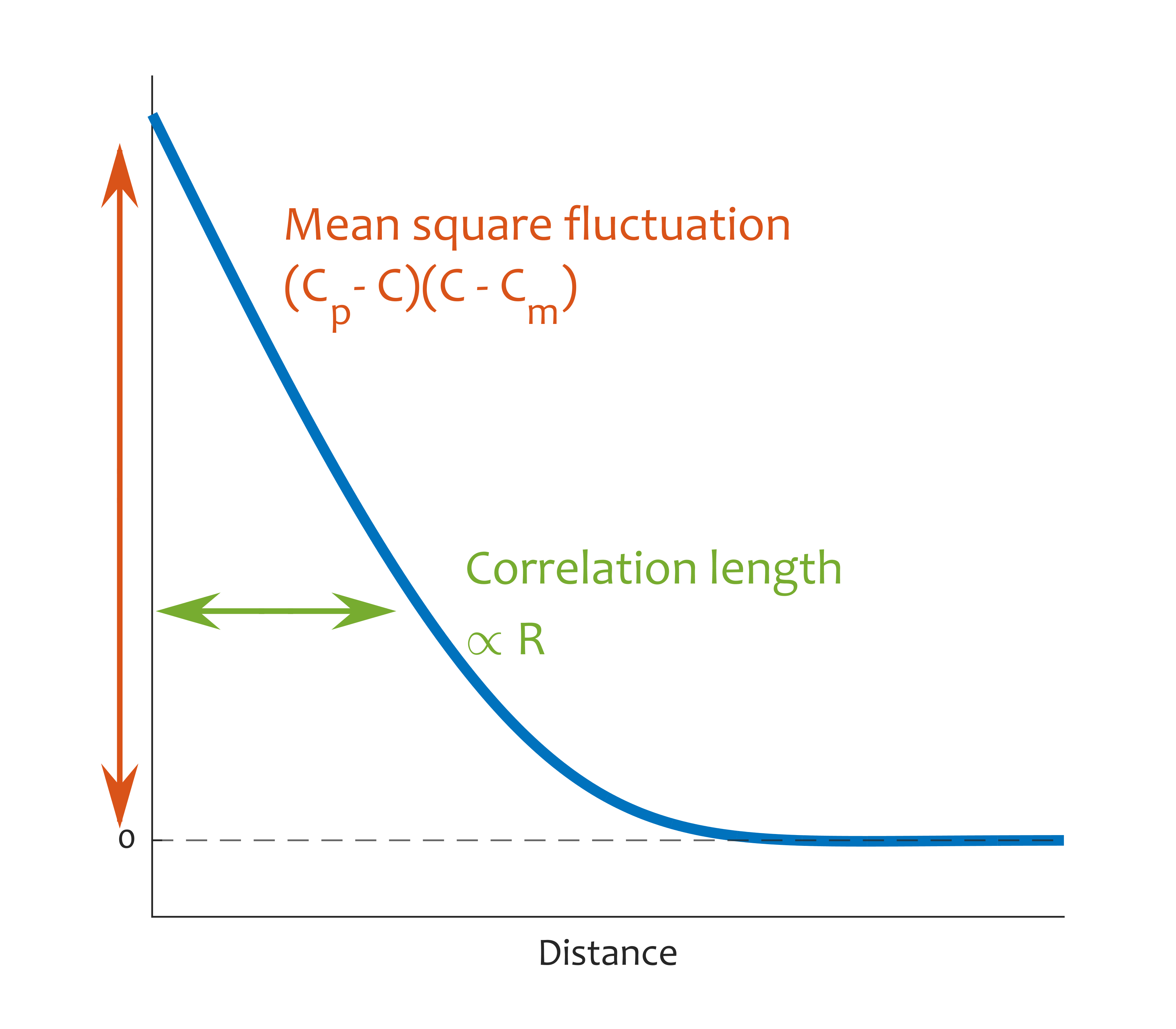}
    \caption{Schematic view of a correlation function in a binary solution. The value at $r=0$ (which should be understood as a limit) is equal to the mean square fluctuation which, in a matrix-precipitates system, equals $(C_p-C)(C-C_m)$. The width of the correlation function is a correlation length, which is related to the size of the fluctuations/precipitates.}
    \label{fig1}
\end{figure}

A cartoon-view of a typical PCF is shown in Fig.~\ref{fig1}. 
Its typical features are highlighted, namely starting from the mean square fluctuation $(C_p-C)(C-C_m)$ and decreasing to zero, with a characteristic correlation length, which is related to the size of the compositional fluctuations, and hence of possible precipitates.
Using the same formalism to process the PCF from APT and the intensity from SAXS allows for directly, and consistently, comparing data from both techniques for characterizing precipitate size, volume fraction, number density for instance. 

\begin{figure}
\centering
\begin{subfigure}{\linewidth}
\centering
    \includegraphics[width=8.5cm]{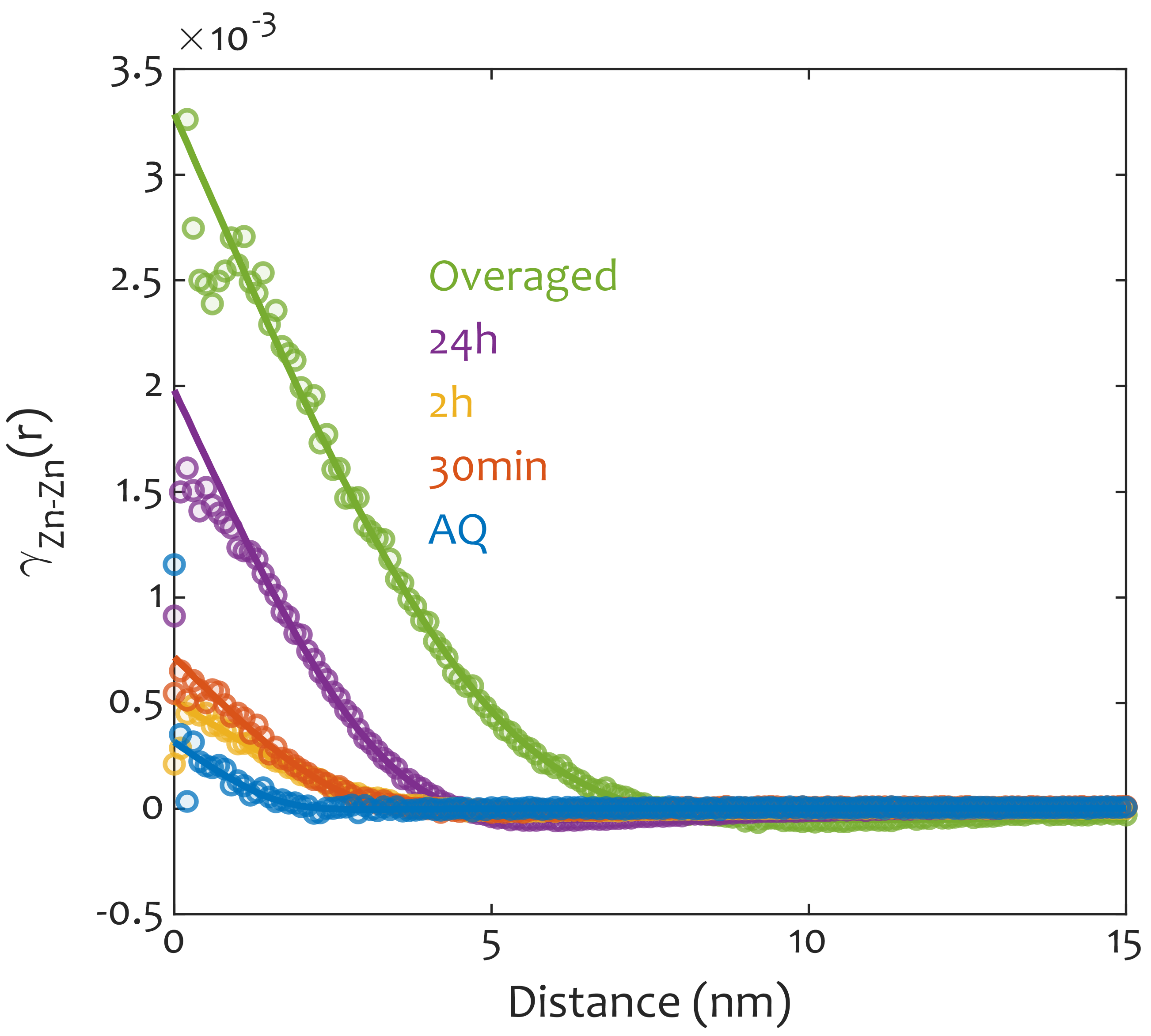}
    \caption{}
    \label{fig2a}
\end{subfigure}

\begin{subfigure}{\linewidth}
\centering
    \includegraphics[width=8.5cm]{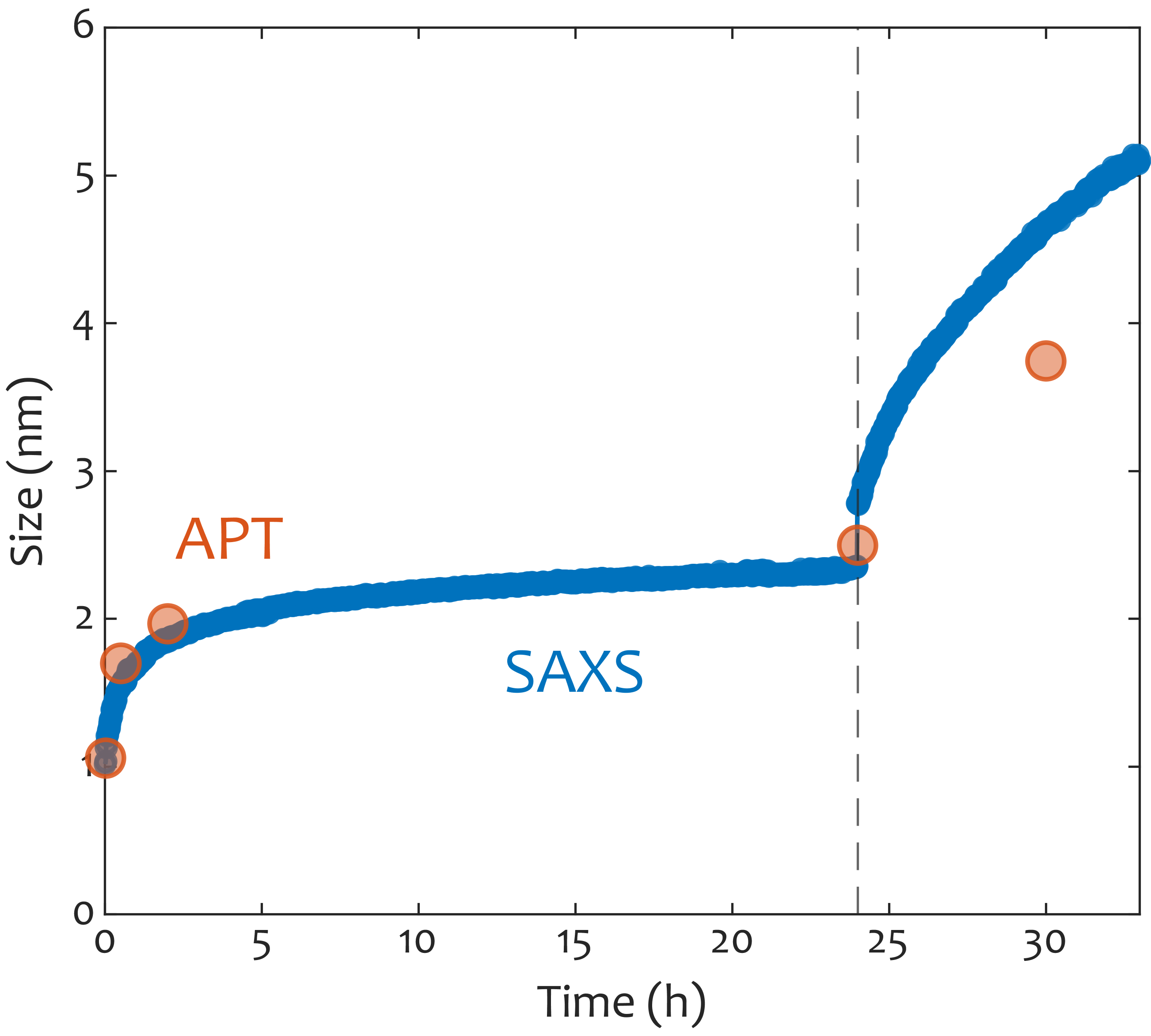}
    \caption{}
    \label{fig2b}
\end{subfigure}
\caption{(a) Zn-Zn pair correlation functions (PCF) obtained by APT on a model Al-Zn-Mg-Cu alloys for different ageing times (circles) along with the fits from the method in \cite{Zhao2018a}. Both the amplitudes and correlation lengths increase with ageing times. (b) Size of the precipitates obtained by \emph{in situ} SAXS (blue dots) and \emph{ex situ} APT (red circles). The agreement is very good.}
\label{fig2}
\end{figure}

The application of the technique is showcased in Fig.~\ref{fig2} across the ageing of a model Al-Zn-Mg-Cu alloy \cite{zhao_segregation_2018}. 
This same alloy was used to introduce the protocol for APT data \cite{Zhao2018a}. 
In Fig.~\ref{fig2a} are the Zn-Zn PCF for different ageing states from as-quenched to overaged, along with the fitted lines coming from the model. 
{\revised In this case, the model was a distribution of spheres. The correlation function of a sphere of radius $R$ is given by \cite{glatter_general_1982}:
\begin{equation}
\gamma_0^\text{sphere}(r)=1-\frac{3r}{4R}+\frac{r^3}{16R}
\end{equation}
when $r<2R$ and $0$ elsewhere. The size distribution can be integrated numerically. Here we used a lognormal distribution with a 20\% dispersity \cite{Zhao2018a}.

The results indicate an increase of both the value at $r=0$ and the correlation length. 
The amplitude increase can be related to an increase of volume fraction and/or composition of the precipitates (eq.~\ref{mean_sq}).
The correlation length increase shows the growth of the precipitates.} 

In Fig.~\ref{fig2b} is the evolution of the radius of precipitates derived from the PCF compared to a SAXS experiment during which the sample was heat-treated \textit{in-situ} at \SI{120}{\celsius} for \SI{24}{\hour}, after which the temperature is raised to \SI{180}{\celsius}, in order to create the overaged state. 
The dashed vertical line marks the change in the temperature of the heat treatment. 
In this case, the match in the particles' radius between the two techniques is particular good, with the notable exception of the last APT point in the \emph{overaged} state for which APT size is smaller. 
This may be explained as the $\eta'$ precipitates are low-evaporation-field particles, and local magnifications make them appear denser and somewhat compressed \citep{vurpillot2000}.

\section{Results and discussion}
\subsection{Precipitate size}
\begin{figure}
    \centering
    \includegraphics[width=8.5cm]{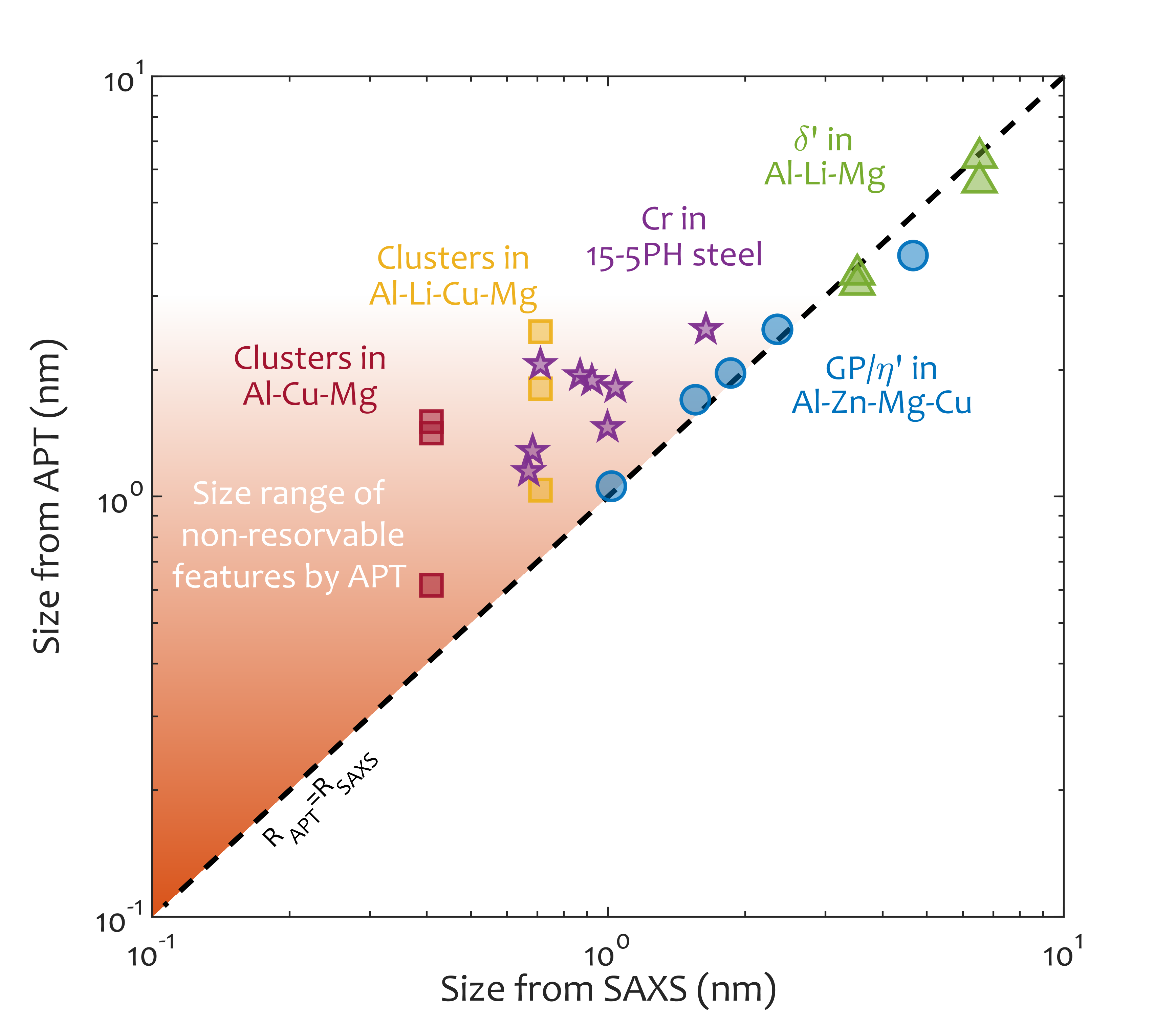}
    \caption{Radius of objects obtained by APT as a function of the radius obtained by SAXS. If we neglect the uncertainty on the size obtained by SAXS, the dashed $R_{APT}=R_{SAXS}$ line should be seen as a reference. The smallest objects all deviate towards larger APT sizes.}
    \label{fig3}
\end{figure}

We have then deployed this approach to a range of different materials, revisiting data where SAXS and APT had been performed on the same alloys, which includes Al$_3$Li precipitation in Al-Mg-Li alloys \cite{Deschamps2012, Gault2012b, de_geuser_complementarity_2014}, Cr decomposition in 15-5PH steels \cite{couturier_direct_2016} and clustering in an Al-Li-Cu-Mg alloy and an Al-Cu-Mg alloy \cite{Ivanov2017}. 
In figure \ref{fig3}, we have represented the radius obtained by APT as a function of the radius obtained by SAXS for these data, along with the results presented in Fig.~\ref{fig2} for the Al-Zn-Mg-Cu alloy. 
The remarkable agreement in the case of Al-Zn-Mg-Cu was already shown in Fig.~\ref{fig2b} and that for Al-Mg-Li had been discussed in an earlier effort \cite{de_geuser_complementarity_2014}. 
The decomposition of Cr in the matrix in 15-5PH steel is however deviating, with the radii obtained from APT about twice larger than those obtained by SAXS. 

The disagreement is even larger for both the Al-Li-Cu-Mg and Al-Cu-Mg alloys where we expect very small clusters. 
{\revised For these two data sets, a single size is reported from SAXS, while 3 distinct sizes have been reported on Fig.~\ref{fig3}. These correspond to the species-specific PCFs for Cu-Cu, Cu-Mg and Mg-Mg, respectively.}

The different composition, and potentially crystal structure, of the particles compared to the matrix leads to differences in the electric field necessary to provoke field evaporation, which is termed evaporation field. 
It is known that a difference in the evaporation field leads to the development of a local radius of curvature at the specimen's surface and hence in the projection's magnification \citep{Miller1987a}. 
In addition to a non-homogeneous magnification, these effects also cause trajectory aberrations and overlaps, leading to the loss of the expected one-to-one mapping between the detector and the specimen's surface. Mathematically, this is called the bijectivity. The significant additional uncertainty as to from which position at the specimen's surface does an ion originate is well documented in the literature \cite{Krishnaswamy1975, Miller1987a}, complemented by results from simulations \citep{vurpillot_model_2013, larson_atom_2013}. 
It is important to bear in mind that the atoms that belong to these particles are detected with a similar efficiency as the atoms from the matrix. This independence of the efficiency against the mass-to-charge-state ratio of the incoming ion is a key strength of APT. 

However, trajectory overlaps will lead to either atoms from the precipitate being imaged within the matrix or vice versa, atoms from the matrix being imaged inside the precipitate. This loss of bijectivity caused by trajectory overlaps is, in many ways, much more problematic than a local change in the magnification in the sense that it involves a stochastic "blurring" of the positions of the atoms. This is highly relevant for precipitates with radii in the range investigated here, i.e. below \SI{10}{\nm}. 

The net effect of this loss of bijectivity is a blurring of the objects so that they appear larger or smaller than they really are, making the data points in Fig.~\ref{fig3} deviate from towards larger apparent size. 
This can be understood as if they were subjected to an effective spatial resolution convoluted with these objects within the volume. 
This effective spatial resolution does not define the smallest detectable object, but the smallest apparent size of an object.

The excellent agreement between APT and SAXS appears for the larger particles, but also in cases where they require a lower evaporation field compared to the matrix. 
This leads to a compression of the trajectories, which, in the case of large particles, leads to an underestimation of the particle size. 
This compression is more markedly visible in the overaged state of the Al-Zn-Mg-Cu alloy, as seen in \ref{fig2b}. 
In the case of small particles, this compression of the trajectories most likely partly compensates the effect of the trajectory overlaps, leading to the excellent agreement of the other data points for Al-Zn-Mg-Cu alloy.
{\revised This is in stark contrast with the results obtained for particles with a higher evaporation field than the matrix \citep{Miller1987a,vurpillot2000}. For instance here, the Cr-rich particles in the 15-5 PH steel or the Cu-rich clusters in the Al-Cu-Mg or Al-Li-Cu-Mg alloys exhibit radius from APT two- to three times larger than those measured by SAXS, with a large scatter in the radius derived from APT across different data sets. This lack in accuracy and precision of the measurement and the discrepancy with SAS will be further discussed below.}

A similar element-specific compression of the trajectories also explains why the sizes obtained on the partial PCFs calculated for same population of clusters are not equal. 
This is the case for the Cu-Cu, Cu-Mg and Mg-Mg PCF from data sets of Al-Cu-Mg and Al-Cu-Li-Mg alloys for instance. 
This observation stresses the fact that the effective spatial resolution of APT should be considered to be element and phase specific, as already pointed out by other authors \citep{Marquis2008f}.

\subsection{Influence of the instrument}
{\revised 
Our results point to some of the key limitations in detecting small particles by APT. 
The data used in Fig.~\ref{fig3} were acquired on a range of different instruments, some with a straight flight path, some fitted with a reflectron, and with different detection efficiencies (see section~\ref{secDisc} for comparisons with even more experimental configurations). 
While many authors discuss the detrimental effect of a limited detection efficiency in the context of clustering studies, based on present findings, we believe that effective spatial resolution is really the first order limiting factor. 
Even in articles claiming otherwise using simulated data (see \cite{Ceguerra2012a} for instance), the influence of the detection efficiency appears minimal.}

{\revised In addition, the data was sometimes acquired with voltage or laser pulsing at various base-temperatures. 
The influence of the base temperature on the precipitate size measurement was shown to be limited \citep{Gasnier2013}, but issues associated to the asymmetrical specimen shapes or variations in the end-radius over the course of the analysis \citep{Gault2010p} are not accounted for in the reconstruction protocol implemented in the commercial software used to process the data \citep{Geiser2009a}. 
These aspects may introduce additional uncertainties are introduced by the experimental conditions, however these have a rather limited effect on the close-neighborhood of each atom \citep{Gault2009e}, and their influence on the detection and size of clusters is hence expected to be limited.
The reflectron may be more critical as it introduces systematic biases in the point density across the field-of-view due to the curvature of the ion trajectories. 
Some are corrected by the analysis software, but how much is difficult to assess. 
All these effects contribute to a worsening of the effective resolution and to the scatter observed in the data in Fig.~\ref{fig6}.

The use of a fixed distance threshold used in the cluster-finding algorithms will lead to cluster detection biases across the field-of-view, i.e. clusters may appear smaller or larger depending on where they are detected because of the reflectron. 
This systematic bias has not been discussed as such. 
The use of an RDF-based technique, which does not use a fixed threshold, allows for circumventing such issues and provides a better average view of the distribution of clusters within the reconstructed volume.}

The strongest limitations is the influence of the trajectory aberrations, which are mostly independent on the design of the instrument, and underpinned by the physics of the field evaporation process.

{\revised \subsection{Influence of molecular ion detection}
Since the effective spatial resolution is dictated by the field evaporation process, an aspect to consider is the possibility that molecular ions can be detected, as is often the case in the analysis of carbon and nitrogen in steels \citep{Sha1992}. Molecular ions are often detected in APT. Since ion trajectories are directly related to the distribution of the electrostatic potential and are independent on the voltage and the ion's mass and charge \citep{Smith1978}, to a first approximation, molecular ions should follow the same trajectories as their atomic counterparts emitted from the same position at the surface of the APT specimen. An added complexity, however, is that molecular ions are often metastable and can dissociate during the flight. Müller et al. reported that such dissociations leads to a difference in the average distance between detector impacts \citep{Muller2011}, which are indicative of additional trajectory aberrations during the ion flight. Further work confirmed these trends in the analysis of oxides and nitrides and carbides \citep{Blum2016, Zanuttini2017, Gault2016, Peng2019}. Depending on the energetic path during these dissociative processes, the possible exchange of energy between the charged fragments that leads to additional aberrations in the ion trajectories that would lead to a further deterioration of the effective spatial resolution \citep{Blum2016,Peng2019}}. 

\subsection{Influence of a limited spatial resolution on the RDF}
The spatial resolution in APT is non-isotropic, with the depth resolution expected to be better the lateral spatial resolution \citep{vurpillot_structural_2001, gault_spatial_2010}. However, the spatial resolution varies systematically across the field-of-view, by a factor of 5 or more across in depth and two or more laterally \citep{Gault2009f}. Changes in resolution with i.e. the base temperature previously reported also mean that during a single analysis in laser pulsing mode, the spatial resolution will also change. Defining an effective resolution that reflects an average resolution is hence a more practical but also more honest way to depict the true performance of the technique. 

In the case of a particle in a matrix, a limited resolution results in an interfacial mixing. Let us consider the case of the Al-Mg-Li alloy. 
High-resolution electron microscopy studies of $L1_2$ ordered precipitates in an Al-matrix, such as the $\delta'$ precipitates in the Al-Mg-Li alloy investigated here, suggest an abrupt interface between precipitate and matrix \cite{schmitzHighResolutionElectron1994}. 
Yet, APT data most often shows a diffuse interface \cite{Gault2012b}, which can be, at least partly, attributed to the limited spatial resolution of the technique. 
Other issues could arise from processing of the data itself \cite{Martin2016}, but they are likely a second order aspect. 
The sharpness of the interface might thus be a first benchmark of the effective spatial resolution.

\begin{figure}
\centering
\begin{subfigure}{\linewidth}
\centering
    \includegraphics[width=8.5cm]{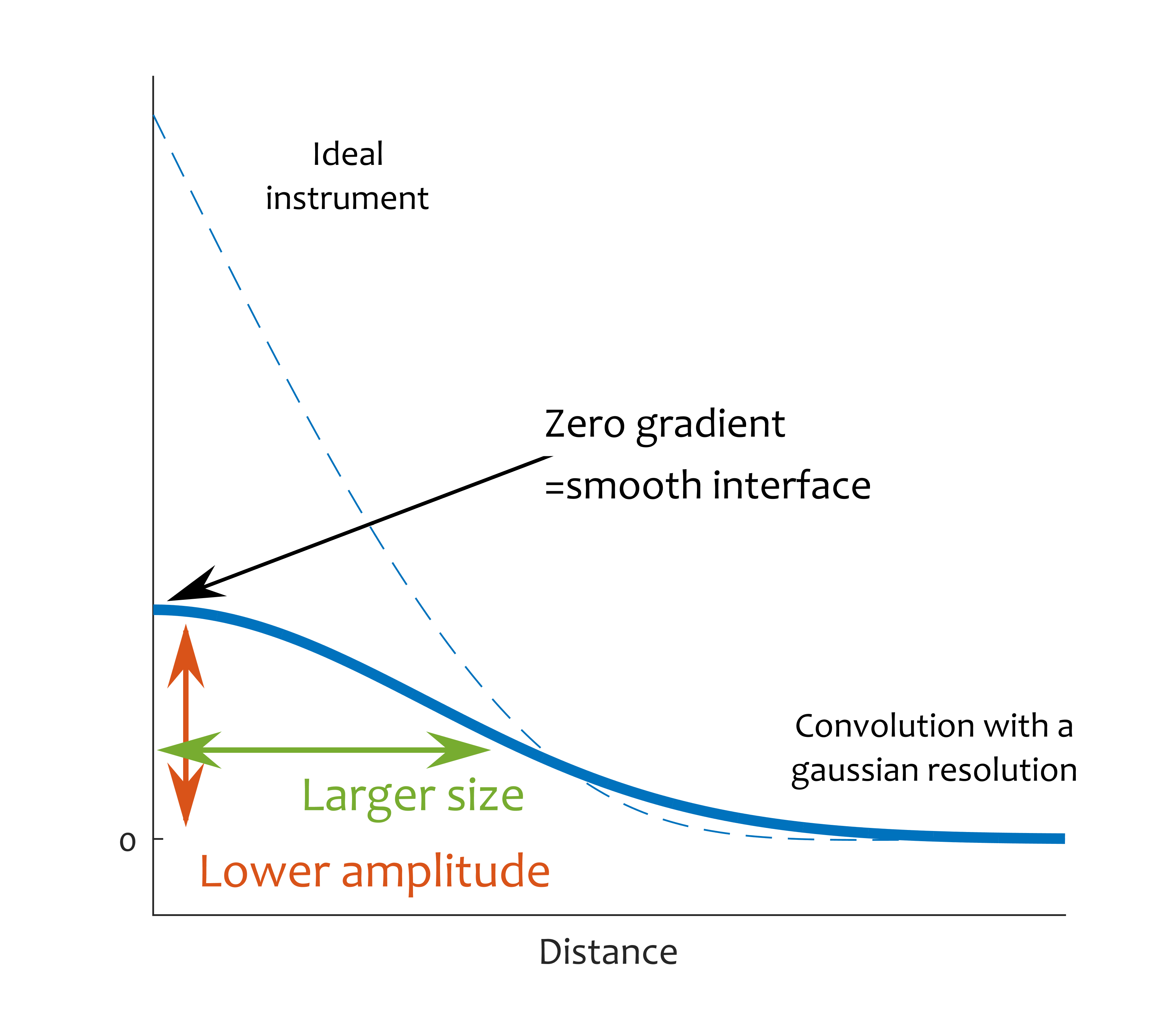}
    \caption{}
    \label{fig4a}
\end{subfigure}

\begin{subfigure}{\linewidth}
\centering
    \includegraphics[width=8.5cm]{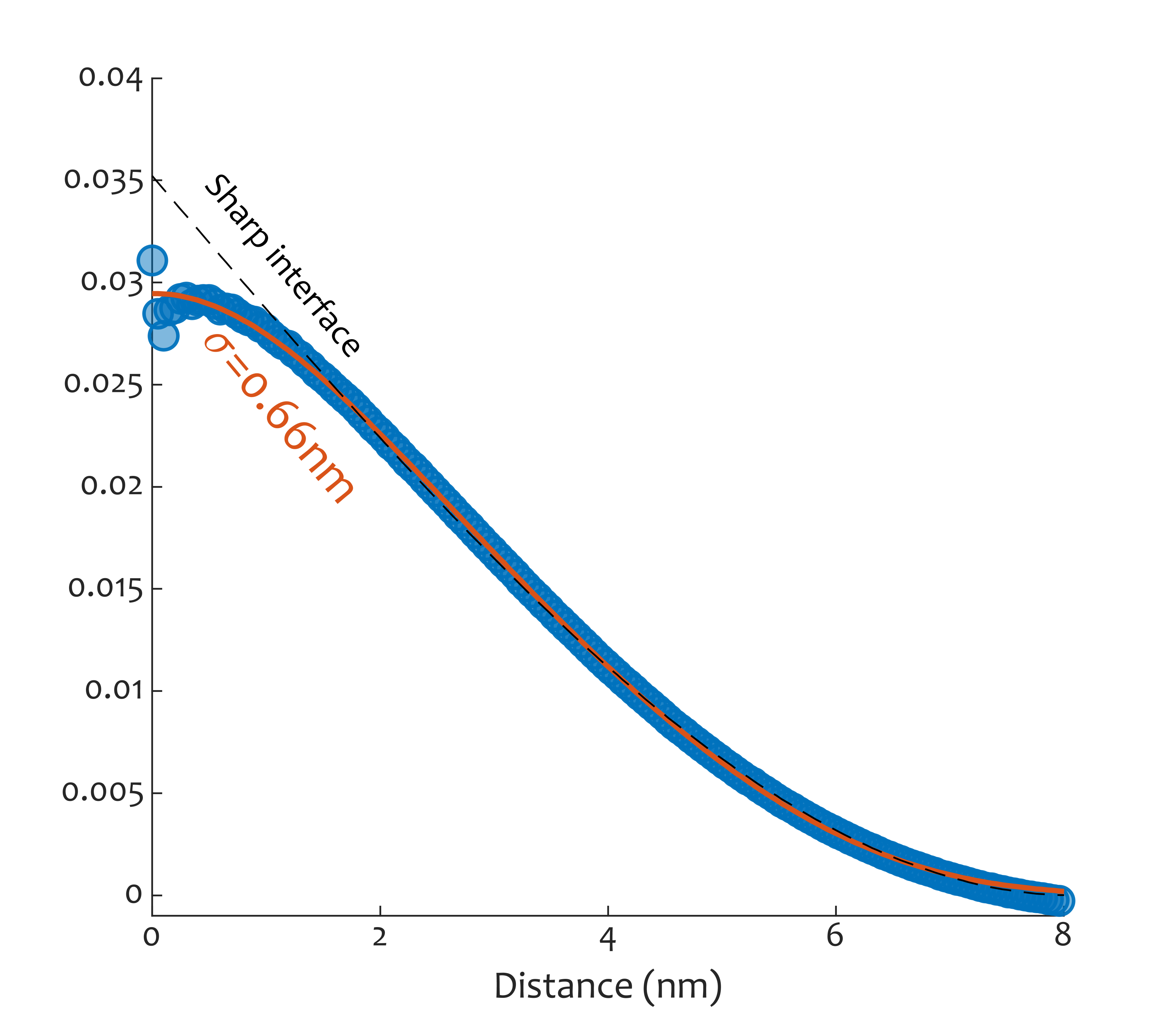}
    \caption{}
    \label{fig4b}
\end{subfigure}

    \caption{(a) Illustration of the effect of the convolution of the PCF with a (Gaussian) spatial resolution. The apparent size is larger, the apparent amplitude smaller and the initial slope indicates a diffuse apparent interface. (b) Experimental Li-Li PCF in an Al-Li-Mg alloy \cite{Gault2012b, de_geuser_complementarity_2014} showing a zero initial slope, consistent with the convolution of a sharp interface with a \SI{0.66}{\nm} spatial resolution.}
    \label{fig4}
\end{figure}

{\revised A limited resolution can be modelled by convoluting the data with a resolution function. }
Since we describe the objects by their pair correlation function (PCF), let us model the effect of the resolution on the shape of the PCF.
We can assume the spatial resolution as being a Gaussian function  with a standard deviation $\sigma$. 
Numerically convoluting a PCF shape by the resolution function leads to a resolution-affected PCF, keeping in mind that the convolution should be performed on $\gamma\cdot r$, since we are in spherical coordinates, see e.g. \cite{egami_underneath_2003}.
The convolution of this Gaussian function with a typical PCF from a precipitate is shown schematically on Fig.~\ref{fig4a}.

An initial important consideration is that the use of a Gaussian function to describe the spatial resolution allows us to quantify the smallest apparent radius that can be measured, i.e. $2\sigma$, since the resulting convoluted PCF will have at least this width.  
Besides a widening of the PCFs, i.e. of the correlation length, the convolution also causes a drop of their amplitude, i.e. of the apparent composition of the particle, which will be discussed further below. 
Another distinctive effect of the convolution with a resolution function is that it tends to flatten the initial slope of the PCF towards zero.
A sharp interface between precipitates and matrix should give rise to a non-zero initial slope in the correlation function. 
A zero initial slope thus corresponds to an apparent diffuse interface, as is visible in Fig.~\ref{fig4a}. 
This is the real space version of what is known in the scattering community as the Porod law \cite{glatter_general_1982}, whereby the initial slope of the correlation function is related to the specific surface, i.e. the inverse of the particle size for spheres.

Let us consider an experimental case. 
As discussed above, the $\delta'$ Al$_3$(Mg,Li) precipitates can be assumed to possess a reasonably sharp interface. Yet, the Li-Li experimental PCF shown in Fig.~\ref{fig4b} shows a distinctive initial slope characteristic of a diffuse interface. 
Here, we were able to reproduce this feature by convoluting the PCF obtained for a sphere by a Gaussian function with a standard deviation of $\sigma_\text{exp}=\SI{0.66}{\nm}$ to account for the spatial resolution. 
This value of $\sigma_\text{exp}$ is extracted from a best fit to the experimental data. 
This effective spatial resolution is relative to the analysis of this particular set of particles and is compatible with the set of observations in Fig.~\ref{fig3} where APT seems to hit a limit in apparent size at about $2\sigma =\SI{1}{\nm}$. 

\begin{figure}
\centering
\begin{subfigure}{\linewidth}
\centering
    \includegraphics[width=8.5cm]{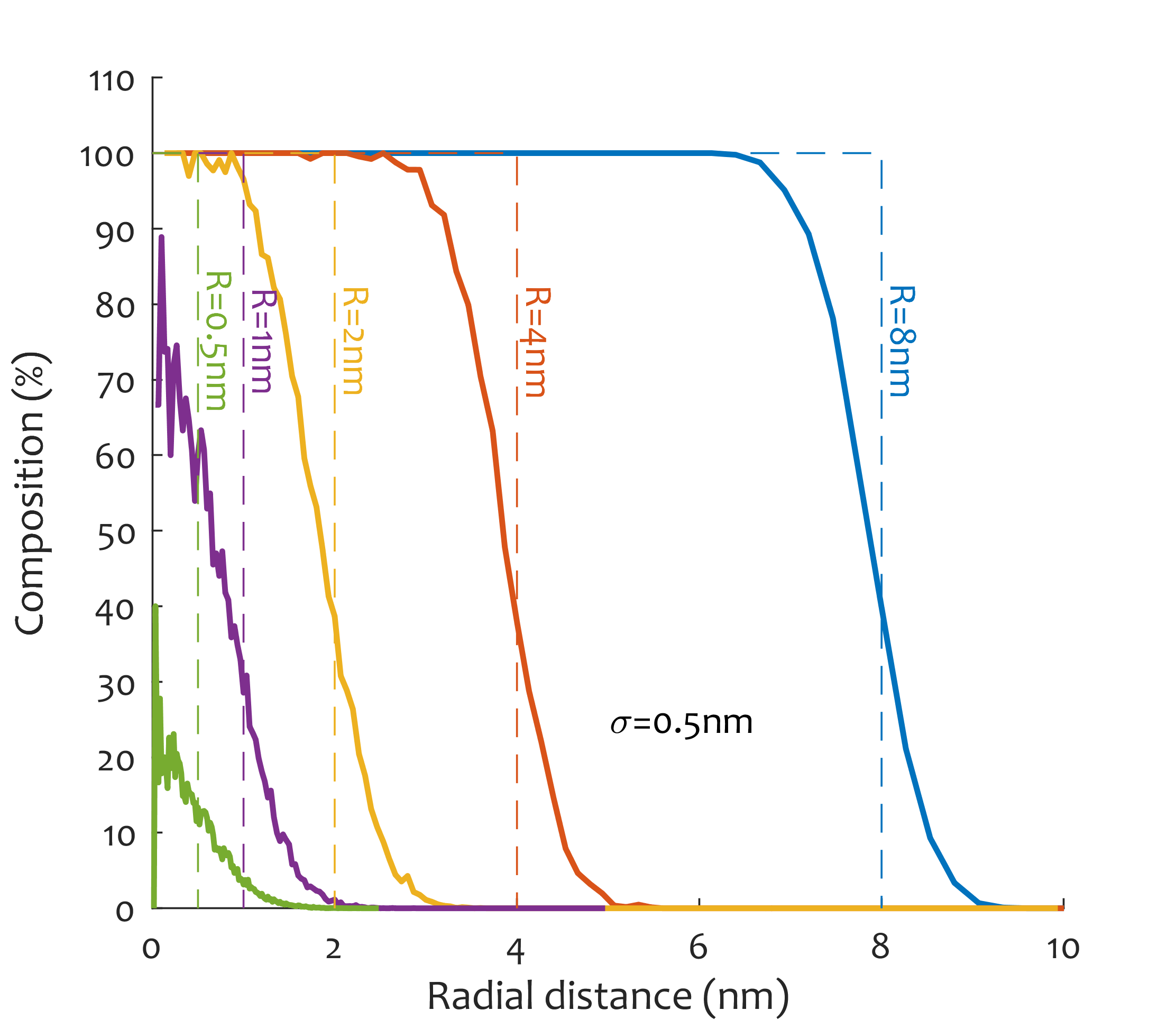}
    \caption{}
    \label{fig5a}
\end{subfigure}

\begin{subfigure}{\linewidth}
\centering
    \includegraphics[width=8.5cm]{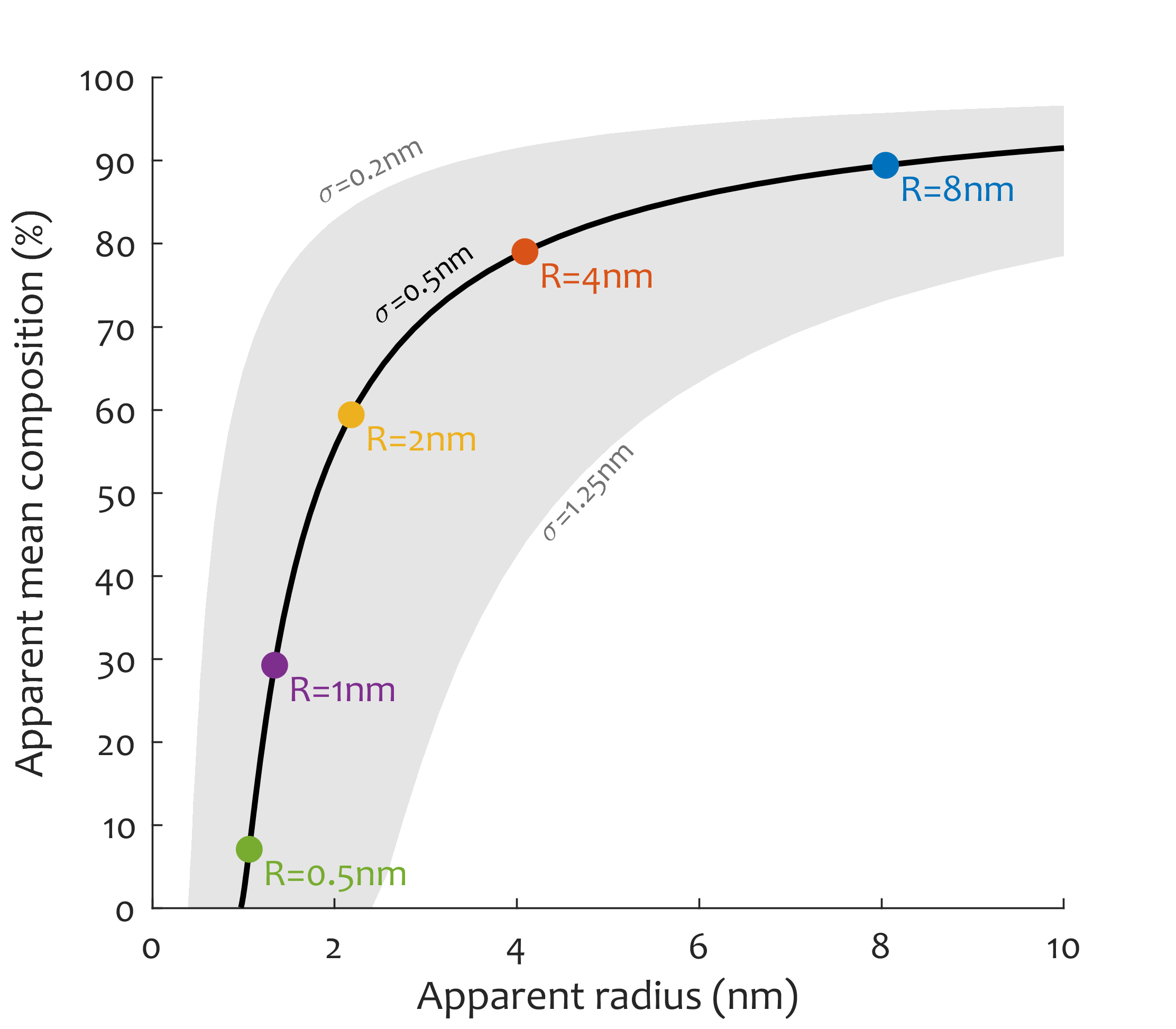}
    \caption{}
    \label{fig5b}
\end{subfigure}
    \caption{(a) Radial composition profiles obtained on simulated data sets containing spherical precipitates of various sizes, with a random noise reproducing a Gaussian spatial resolution of $\sigma=\SI{0.5}{\nm}$. For smaller sizes, even the core composition is affected. (b) Mean composition of the precipitates above, showing that a cluster identification will significantly underestimate the solute content of precipitates, even for relatively large sizes. The solid line was obtained separately by computing the drop in amplitude of an analytical spherical PCF due to convolution with a Gaussian of $\sigma=\SI{0.5}{\nm}$, confirming that the PCf formalism is affected the same way. The grayed area is bounded by the curves obtained for $\sigma=\SI{0.2}{\nm}$ and $\sigma=\SI{1.25}{\nm}$.}
    \label{fig5}
\end{figure}

Beside the effect on the apparent size and on the apparent sharpness of the interface, we can now evaluate the effect of this spatial resolution of the apparent composition of the particles detected by APT.
To better illustrate this influence, we generated a series of virtual data sets containing a spherical particle with a composition of 100\% of solute in a matrix devoid of solutes. 
The radius of the precipitate was varied in the range of \SIrange{0.5}{8}{\nm}. 
To simulate finite spatial resolution, a Gaussian noise with $\sigma=\SI{0.5}{\nm}$ is used to randomly shift the position of each atom. 
We plot radial composition profiles obtained on each virtual data set in Fig.\ref{fig5a}.
A radial composition profile is an ideal situation in that the particles are isotropic by construction and we know the position of their centre.
This emulates the effect of the resolution as measured by a 1D profile or a proxigram, suppressing any smoothing effects due to the data processing itself \citep{Hellman2003, Martin2016}.

For larger precipitates, only an intermixing of the particle and the matrix appears near the interface.  
Below \SI{2}{\nm}, however,  the limited resolution causes a significant drop in the composition of the particle even at the core of the particle. 
The apparent composition drops from 100\% down to \textapprox25\% for the smallest precipitate size. 

While this gives an estimate of the effect of a spatial resolution on one-dimensional profiles, it is also interesting to assess how it would affect cluster identification method results on the precipitate contents. 
Let us assume an optimized cluster identification methods which correctly identifies all solutes atoms belonging to the precipitates. 
The average composition of the detected cluster could then computed from the profiles in Fig.~\ref{fig5a}, where the average is weighted by the number of atoms in each class. 
The results are shown in figure \ref{fig5b}, where the average composition of each simulated precipitate is represented by the colored circles. The grayed area corresponds to the range covered by the function for values of $\sigma$ ranging from \SIrange{0.2}{1.5}{\nm}. 

The solid line in Fig~\ref{fig5b} has been obtained by convoluting an analytical PCF with a Gaussian function and plotting the drop in amplitude, confirming that the PCF formalism is affected by the same bias than the average composition.
We can see that the mean composition is always significantly below 100\%, even for precipitates as large as \SI{8}{\nm} in radius. 
It even drops below 10\% for the \SI{0.5}{\nm} particle.
It is counter-intuitive that even larger particles are so much affected, since blurring should affect only the interfacial area.
The reason for this is that, most of the atoms are, in fact, within the interfacial region. 

\subsection{Comparison with experimental reports from the literature}
\label{secDisc} 
\begin{figure*}
    \centering
    \includegraphics[width=12cm]{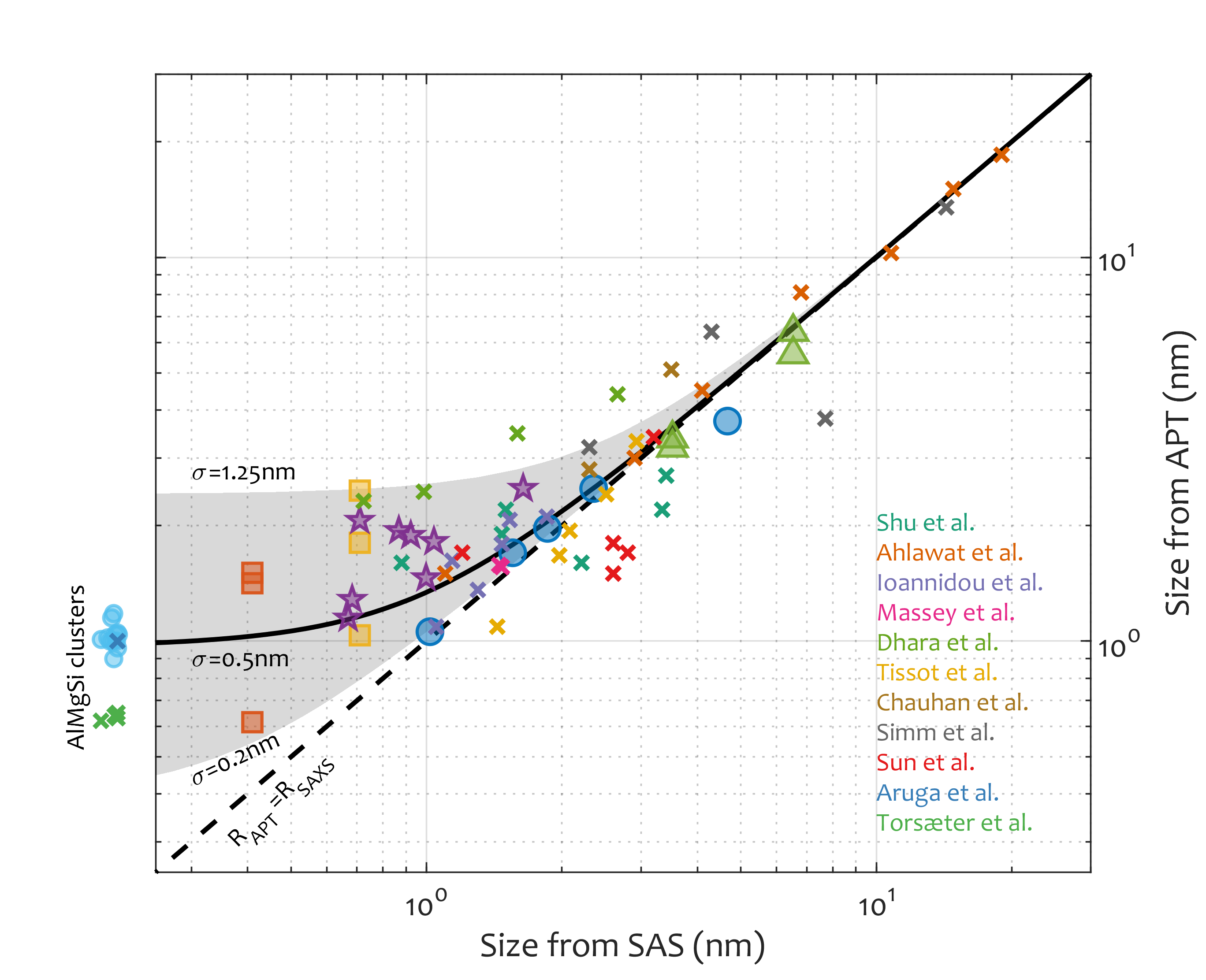}
    \caption{Radius of objects obtained by APT as a function of the radius obtained by small-angle scattering (SANS or SAXS). 
    The solid line is the modeled effect of a Gaussian resolution function with  $\sigma=\SI{0.5}{\nm}$, along with a grayed area bound by the $\sigma=\SI{1.25}{\nm}$ and $\sigma=\SI{0.2}{\nm}$ curves.
    In addition to the data points from Fig.~\ref{fig3}, we have extracted data from recent literature as crosses \cite{shu_multi-technique_2018, ahlawatRevisitingTemporalEvolution2019, ioannidouInteractionPrecipitationAustenitetoferrite2019, masseyMultiscaleInvestigationsNanoprecipitate2019, dharaPrecipitationClusteringTiMo2018, tissotComparisonSANSAPT2019, chauhanMicrostructureCharacterizationStrengthening2017, simmSANSAPTStudy2017, sun_g_2017, aruga_formation_2015, torsaeterInfluenceCompositionNatural2010}. The symbols outside the axis correspond to clusters in Al-Mg-Si alloys where SAS sizes are unavailable. Clusters in these alloys are expected to be very small.}
    \label{fig6}
\end{figure*}

Regarding considerations of particle size, we plotted in Fig.~\ref{fig6} the apparent size of the particles as a function of their real size as reported by experimental studies using both SAXS or SANS and APT. 
The crosses correspond to data from recent literature \citep{shu_multi-technique_2018, ahlawatRevisitingTemporalEvolution2019, ioannidouInteractionPrecipitationAustenitetoferrite2019, masseyMultiscaleInvestigationsNanoprecipitate2019, dharaPrecipitationClusteringTiMo2018, tissotComparisonSANSAPT2019, chauhanMicrostructureCharacterizationStrengthening2017, simmSANSAPTStudy2017, sun_g_2017, aruga_formation_2015, torsaeterInfluenceCompositionNatural2010} covering a range of alloy systems, including maraging steels, Fe-Cu alloys, ODS-steels, Al-based alloys. 
We simply plot the values reported in tables or graphs. Please note that the values of the size in SAS were obtained through different data processing routines. 
The other symbols correspond to data already discussed above. 
The data points to the left of the y-axis are for Al-Mg-Si alloys for which SAXS does not yield satisfactory signal, and only APT data is available (see below).

Superimposed as a black line is the result of the convolution of an analytical PCF with a Gaussian function, and the resulting apparent size obtained for $\sigma=\SI{0.5}{\nm}$. 
The grayed area is bound by the curves corresponding to $\sigma=\SI{1.25}{\nm}$ and $\sigma=\SI{0.2}{\nm}$.
Figure \ref{fig6} confirms that experimental sizes obtained by APT are compatible with an effective size resolution in the range $\sigma=\SI{0.5}{\nm}$, which does not allow an apparent size below $2\sigma$, i.e. below \textapprox\SI{1}{\nm}. 

This limit does not correspond to the smallest detectable objects, but to the precision limit, i.e. the smallest apparent size APT is able to report. This corresponds to the point spread function of the instrument, i.e. its response to a sub-resolution object. It is crucial to remember that this might be specific to an alloy system, its microstructure and the analysis condition, as can be inferred from the scatter of the data which is reasonably captured by the grayed area.

\subsection{Comparison with spatial resolutions considered in the literature}
These values of $\sigma$ are much broader than spatial resolution values usually reported for APT in the literature. 
For better comparison, it is important to keep in mind that $\sigma$ is here the standard deviation of an isotropic Gaussian resolution function. 
The literature often quote the resolution as $2\sigma$ or FWHM ($2.35\sigma$) which, for the values of $\sigma$ reported here, would put the resolution around \SI{1}{\nm} with a minimum of \SI{0.5}{\nm} and values as high as \SI{3}{\nm} in many cases.

In addition, the effective spatial resolution should be considered as a rotational average of the resolution in all directions. 
Since APT resolution is considered anisotropic, some APT resolution estimations report separated values for depth resolution $\sigma_z$ and lateral resolution $\sigma_{xy}$. In this case, the average effective resolution can be expressed as 
\begin{equation}
    \sigma_\text{eff}=\sqrt{\frac{2\sigma_{xy}^2+\sigma_z^2}{3}}\label{eqSigma}
\end{equation}

Several spatial resolution values for APT can be found in the literature (e.g. \cite{vurpillot_structural_2001, Cadel2009, gault_spatial_2010, geiserSpatialDistributionMaps2007, haley_influence_2009}) which, for $2\sigma$, are in the range \SI{0.05}{\nm} in depth and about \SI{0.2}{\nm} laterally in the worse cases. 
This corresponds to an effective $\sigma$ of \SI{0.08}{\nm} where what we report is up to 15 times this value. 
This should clearly be attributed to the fact that spatial resolution of APT has traditionally been estimated in ideal situations with little practical interests.

\section{General discussions}
With APT being sometimes referred to as a microscopy technique, the question of the resolution very often arises.
Existing attempts at estimating the resolution have focused on the highest achievable spatial resolution, obtained on atomic planes in a limited sub-volume along a given crystallographic direction.
The value of this optimal spatial resolution is, however, only marginally useful since it corresponds to situations of virtually no practical interest. The results shown here are an attempt to estimate an effective spatial resolution in a practical context, namely the study of clusters/precipitates in a matrix.

While many authors discuss the detrimental effect of a limited detection efficiency in the context of clustering studies, based on present findings, we believe that effective spatial resolution is really the first order limiting factor. Even in articles claiming otherwise using simulated data (see \cite{Ceguerra2012a} for instance), the influence of the detection efficiency appears minimal. 

However, up to now, there was a lack of benchmark against which APT sizes could indeed be measured or estimated. Comparisons with other techniques were reported for features of a size that is large enough to be measured by transmission electron microscopy \cite{Gasnier2013} or secondary-ion mass spectrometry have been reported \cite{Ronsheim2008}. 
However, the match was often poor for smaller features like, for e.g., solute clusters, which remain physically ill-defined and for which there are no reference from real-space analytical techniques. 
Here, we used information gathered from SAS. 
When the signal is high enough, SAS can provide accurate size measurement for extremely small particles. 
{\revised A lower limit is sometimes reported in the literature (e.g. \textapprox\SI{1}{\nm} \citep{Hyde2014} for SANS) but it relates to an instrument-dependent estimation where the signal becomes too low to be detected. 
For SAS, the key issue is the signal intensity, and not the spatial resolution: if a signal is gathered, it is precise.}

However, since SAXS relies on electronic density contrast and SANS on scattering length density, there are cases where SAS is essentially blind to clusters. It is for instance the case of the Al-Mg-Si alloys which have been subject to numerous studies since their mechanical properties of industrial relevance are due to very fine clusters which prove not resolvable by any other technique than APT \citep{Edwards1998}.

Since it is a classic system for clustering studies, we have added to Fig~\ref{fig6} the size obtained by APT PCF on several Al-Mg-Si-Cu alloys in different states in the course of an unpublished study.
In total, they represent 12 different metallurgical states.
Since we have no SAXS results to benchmark the values, and since we expect them to be small, we represented the 12 data points outside the graph, to the left of the ordinate axis.
The important result is that they are all about \SI{1}{\nm} in radius. The physical size of the clusters is not often reported in the literature, but when it is, a value of about \SI{1}{\nm} is often found and we have added results from this alloy system as crosses on the left of the axis \cite{torsaeterInfluenceCompositionNatural2010, aruga_formation_2015}.
In many of the AlMgSi clustering studies, a radius of \SI{1}{\nm} is actually the smallest detected size. 
A sphere of radius \SI{1}{\nm} represents 250 to 300 atoms, depending on the atomic volume, which is a surprisingly high value for a minimal size of detected clusters.

The point spread function of APT can not be considered as a unique feature, as is possibly the case for some microscopy techniques. 
It is directly related to how trajectory aberrations will affect the measurement, and this is directly dependent on the size and composition of each features of interest. 
In addition, it is affected by how the measurement itself is performed \cite{Miller1981a,Yao2011a}, by how the raw data is reconstructed \cite{Gault2009e,Gault2010p} and, in many cases, by the selection of parameters used to extract information from the reconstructed data \cite{Marceau2011}. So, again, establishing a single value of the spatial resolution or a unique definition of a point spread function is likely impossible. 

Nevertheless, we show here that in the context of clustering/precipitation studies, the effective point spread function is relatively well represented by a Gaussian blur of about \SI{0.5}{\nm} but could possibly be as large as \SI{1.25}{\nm} or more.
These values should be compared to typical nearest neighbours distances on a crystalline lattice. 
They correspond typically to distances above the 5th nearest neighbour distance.
The original vicinity and nearest neighbours on the lattice can not be hoped to be retrieved, which should be considered when attempting e.g. lattice rectification \cite{moodyLatticeRectificationAtom2011, gault_nexus_2017} or when computing short range order parameters from APT data \cite{Ceguerra2012a}.
This means simply that, except maybe very locally for very specific cases, all information on the local neighborhood of the atoms is lost and, generally speaking, so is the crystalline nature of the sample. Only traces might remain.
In fact, Ceguerra \emph{et al.} tried to estimate the effect of an imperfect resolution on short range order from APT, but used values which we now know are unrealistically small ($\sigma_\text{eff}$ of at most \SI{0.12}{\nm}) \cite{Ceguerra2012a}.

Overall, there is a need for the APT community to be more considerate when reporting size and composition of particles with radii below \textapprox\SIrange{5}{10}{\nm}.
The visual aspects of APT are both a blessing and a curse – practitioners want to see a population of clusters and have developed and applied typical clustering algorithms. 

In addition, a value of $\sigma=\SI{0.5}{\nm}$ is typically in the range of the $d_\text{max}$ distances used as a threshold in cluster identifications techniques (e.g. maximum separation or other related). When affected by trajectory aberrations, the identification of clusters based on a threshold of distance or composition \cite{Dumitraschkewitz2018} is hence likely to fail to assess which atoms physically belong to clusters. While several studies have investigated the effect of an imperfect reconstruction (by adding random Gaussian noise to a synthetic APT volume) on cluster identification methods, most concluded that it did not yield too strong an effect on the clusters on the basis of a strongly underestimated effective spatial resolution. 
For instance, Hyde \emph{et al.} \cite{hyde_sensitivity_2011} used what can be computed through eq.~\ref{eqSigma} as an effective resolution of \SI{0.19}{\nm}, Ghamarian \emph{et al.} \cite{ghamarianHierarchicalDensitybasedCluster2019} used \SI{0.35}{\nm} and \SI{0.52}{\nm} (and concluded that clusters below \SI{1}{\nm} were not correctly measured), Hatzoglou \emph{et al.} \cite{hatzoglouQuantificationAPTPhysical2019} used \SI{0.16}{\nm} and \SI{0.39}{\nm} (and started their analysis on \SI{1}{\nm} precipitates), Hyde \emph{et al.} \cite{hyde_analysis_2017} used  \SI{0.25}{\nm} and \SI{0.5}{\nm} and Jägle \emph{et al.} \cite{jagleMaximumSeparationCluster2014} used \SI{0.2}{\nm}.

The values that we report here seem to correspond to the worst case scenario of these studies, when problems start to arise. 
Possibly, most of the authors expected that the detection efficiency would be the worst offender in terms of cluster identification, while most studies show that is has only marginal effects (see e.g. \cite{jagleMaximumSeparationCluster2014}). More worryingly, applying such methods could result in the creation of ``ghost'' clusters, in particular in the case where the matrix contains significant amount of solutes.  
Further investigations are required on these aspects, supported by field evaporation simulations.

We show that the lack of precision in the measurement and the complexity of the relationship between composition, aberrations and measurement accuracy make RDF-based analyses more robust to characterise a population of particles. When it is difficult do find an unambiguous definition of the limits between the objects and their surrounding, one should abandon the idea of imaging each individual cluster and rather rely on a statistical evaluation of the compositional fluctuations. 

The measurement of the composition is obviously made difficult if not impossible by the trajectory overlaps and associated blurring of the positions, as demonstrated by Fig.~\ref{fig5b}.
Our simulations were performed in an oversimplified case of 100\% solute precipitates in a 0\% solute matrix.
The case of less concentrated and more diffuse clusters in a matrix containing solute is clearly more complex, indicating that this situation might lead to results deviating even more from reality.
Multi components system, might particularly suffer from this effect. 
Even in the case where only the solute content obtained from cluster identification methods is considered, the ratio between the different elements may be more strongly affected than expected if the matrix also contains solutes.
Some efforts to offer corrections have been pursued \cite{Blavette2001, Philippe2010} but are clearly not applicable to particles below \SI{1}{\nm} where defining a local density is not trivial, and might not be able to correct for the influence of aberrations.  

We show that composition, volume fraction and sizes are very much affected by the resolution. 
This can be seen clearly in Fig.~\ref{fig4a} where the convolution of the PCF with the resolution is shown to decrease amplitude and increase size.
Using the amplitude of the RDF alone, as is sometimes done (see for instance ref. \cite{Zhou2013a}), is subject to the same bias and will lead to inaccurate compositions since it is very much influenced by the convolution with the spatial resolution. Comparison of the amplitude between microstructural states should only be performed with great care since the resolution varies across data sets. 
However, the conservation of matter imposes that the integral $\int4\pi\gamma r^2 \mathrm{d}r$ be unaffected by the convolution.
This integral corresponds to a mean square number of excess solutes, which could be a good alternative measure of the advancement of a clustering reaction, as proposed by Ivanov \emph{et al.} \cite{Ivanov2017} in the context of small-angle scattering.

The message of this work is not that small features are not detected by APT. 
What we have evidenced is that the width of the point spread function might be wider than what most user would expect, so that the image of small features may be vastly deformed when imaged by APT. 
It is important to acknowledge this, so that the community can work on reliable metrics which are not too dependent on the spatial resolution. 

\section{Conclusion}
{\revised To conclude, we have demonstrated that the effective spatial resolution of APT in the context of the metrology of small object is {\revised worse} than often reported, which will particularly impact the measurement of solute clusters and small precipitates. The size of the PSF is indeed larger than the usually quoted values corresponding to near-ideal situations.}
We have discussed the fact that when approaching \SI{1}{\nm} in radius, the measured values of size and composition of particles by APT should be considered highly questionable. This was enabled by comparison of results with SAS performed on the same materials, as well as by the use of a common framework to process the data. 
Our results demonstrate that, while APT is the only technique capable of analysing clusters of a few atoms, its results should be taken with caution.
We also point towards possible routes for more reliable results on very small objects, namely using statistical methods such as RDF based analyses and including the effect of the point spread function in the interpretation of these analyses.

\section*{Acknowledgements}
The authors would like to thank Dr H. Zhao for providing the samples used in the in-situ SAXS experiment and kindly sharing her experimental APT data allowing for direct comparison. 

\bibliographystyle{elsarticle-num}
\bibliography{roger}

\end{document}